\documentclass[%
reprint,
 superscriptaddress,
 bibnotes,
 amsmath,amssymb,
 aps,
 pra,
]{revtex4-2}

\usepackage{graphicx}
\usepackage[caption=false]{subfig}
\usepackage{qcircuit}
\usepackage{dcolumn}
\usepackage{bm}
\usepackage{tikz}
\usetikzlibrary{shapes}
\usetikzlibrary{shapes}

\usetikzlibrary{plotmarks}
\usetikzlibrary{decorations.markings}
\usepackage{siunitx}
\usepackage{xy}
\usepackage{dsfont}
\usepackage{pgfplots}
\usepackage{pgf}
\pgfplotsset{compat=1.15}
\usepackage{placeins}
\usepackage{flafter}
\usepackage{color}
\usepackage{orcidlink}
\usepackage{hyperref}
\hypersetup{
	colorlinks=true,
	linkcolor=blue,
	filecolor=blue,      
	urlcolor=blue,
	citecolor=blue,
	pdftex,
	pdfauthor={Ahmed M. Farouk, I. I. Beterov, P. Xu and I.I. Ryabtsev},
	pdftitle={Scalable Heteronuclear Architecture of Neutral Atoms Based on EIT},
   pdfcreator={Ahmed M. Farouk},
}
\usepackage{epstopdf}
\makeatletter
\newcommand\tinyv{\@setfontsize\tinyv{5pt}{5}}
\makeatother

\usepackage[mode=buildnew]{standalone}
\usepackage{tikz}

\usetikzlibrary{arrows}

\usetikzlibrary{shadows.blur}
\makeatletter
\tikzset{render blur shadow/.code=
	{\pgfbs@savebb
		\pgfsyssoftpath@getcurrentpath{\pgfbs@input@path}%
		\pgfbs@compute@shadow@bbox
		\pgfbs@process@rounding{\pgfbs@input@path}{\pgfbs@fadepath}%
		\pgfbs@apply@canvas@transform
		\colorlet{pstb@shadow@color}{teal!\pgfbs@opacity!black}%
		\pgfdeclarefading{shadowfading}{\pgfbs@paint@fading}%
		\pgfsetfillcolor{black}%
		\pgfsetfading{shadowfading}%
		{\pgftransformshift{\pgfpoint{\pgfbs@midx}{\pgfbs@midy}}}%
		\pgfbs@usebbox{fill}%
		\pgfbs@restorebb
	},
}
\makeatother

\begin{document}

\preprint{APS/123-QED}

\title{Scalable Heteronuclear Architecture of Neutral Atoms Based on EIT} 

\author{Ahmed M. Farouk \orcidlink{0000-0002-6230-1234}}
\email[]{ahmed.farouk@azhar.edu.eg}
\affiliation{Novosibirsk State University, 630090 Novosibirsk, Russia}
\affiliation{Faculty of Science, Al-Azhar University, 11884, Cairo, Egypt}
\affiliation{Rzhanov Institute of Semiconductor Physics SB RAS, 630090 Novosibirsk, Russia}
\author{I.I. Beterov \orcidlink{0000-0002-6596-6741}}
\email[]{beterov@isp.nsc.ru}
\affiliation{Novosibirsk State University, 630090 Novosibirsk, Russia}
\affiliation{Rzhanov Institute of Semiconductor Physics SB RAS, 630090 Novosibirsk, Russia}
\affiliation{Institute of Laser Physics SB RAS, 630090 Novosibirsk, Russia}
\affiliation{Novosibirsk State Technical University, 630073 Novosibirsk, Russia}

\author{Peng Xu\orcidlink{0000-0001-8477-1643}}
\affiliation{State Key Laboratory of Magnetic Resonance and Atomic and Molecular Physics, Innovation Academy for Precision Measurement Science and Technology, Chinese Academy of Sciences, Wuhan 430071, China}
\affiliation{Wuhan Institute of Quantum Technology, Wuhan 430206, China}


\author{I.I. Ryabtsev}
\affiliation{Novosibirsk State University, 630090 Novosibirsk, Russia}
\affiliation{Rzhanov Institute of Semiconductor Physics SB RAS, 630090 Novosibirsk, Russia}

\date{\today}
\begin{abstract}
Based on our recent paper [\href{https://arxiv.org/abs/2206.12176}{arXiv:2206.12176 (2022)}], we propose a scalable heteronuclear architecture of parallel implementation of CNOT gates in arrays of alkali-metal neutral atoms for quantum information processing. We considered a scheme where we perform CNOT gates in a parallel manner within the array, while they are performed sequentially between the pairs of neighboring qubits by coherently transporting an array of  atoms of one atomic species (ancilla qubits) using an array of mobile optical dipole traps generated by a 2D acousto-optic deflector (AOD). The  atoms of the second atomic species (data qubits) are kept in the array of static optical dipole traps generated by spatial light modulator (SLM). The moving ancillas remain in the superposition of their logical ground states without loss of coherence, while their transportation paths avoid overlaps with the spatial positions of data atoms. We numerically optimized the system parameters to achieve the fidelity for parallelly implemented CNOT gates around $\mathcal{F}=95\%$ for the experimentally feasible conditions. Our design can be useful implementation of surface codes for quantum error correction. Renyi entropy and mutual information are also investigated to characterize the gate performance.
\end{abstract}

\maketitle

\section{Introduction}

Quantum computers have been progressing remarkably in the last few years. This included substantial progress in developing the quantum processors or quantum simulators based on alkali-metal neutral atoms \cite{ebadi2021quantum, scholl2021quantum, graham2022multi} and alkaline-earth atoms \cite{madjarov2020high}. The approaches of demonstrating a quantum processor with dynamic connectivity of atoms have attracted researchers \cite{hansel2001magnetic, beugnon2007two, hickman2020speed, bluvstein2022quantum}. These approaches are essential for building scalable quantum information systems. In Ref.~\cite{hansel2001magnetic}, an integrated magnetic device is used to transport cold atoms near a surface with very high positioning accuracy while the atoms being confined in all three dimensions. However,  high velocity of atomic motion results in heating of the transported atoms. Using optical tweezers it was possible to coherently transport atomic qubits without loss of coherence, as it has been demonstrated in \cite{beugnon2007two}. It has been shown that it is possible to move an atom between two traps having the same depth and it has been found that the transfer using optical tweezers does not induce any significant motional heating.

Bluvstein et al. \cite{bluvstein2022quantum}, demonstrated experimentally the possibility to implement quantum gates and perform error-correction codes with dynamic reconfiguration, when atoms are prepared into two sets of traps. The static traps were generated using a spatial light modulator (SLM), which can directly generate  trapping arrays with arbitrary spatial configuration. The acousto-optic deflector (AOD), which can form regularly shaped atomic arrays when it is driven by a multi-frequency RF field, was used to create mobile traps.  Crossing of trapping positions between different arrays could be avoided. The dynamic configuration of the atomic arrays can be used for storage and transport of quantum information in between quantum gates. The excitation of atoms into Rydberg states is required for generating quantum entanglement. These ingredients enable a powerful quantum information architecture. In the homonuclear configurations of the atomic array, when all atoms are of identical species, at the beginning the atoms are loaded in the static traps, and then the  rearrangement of atoms is performed to create a defect-free array. Then some of the atoms, selected as ancillas, are transfered to the mobile traps \cite{ebadi2021quantum}. The ancillas can be transported into the traps created by AOD, since it allows faster transportation of atoms compared to what is possible with the SLM. 

Heteronuclear architectures can be advantageous due to the additional control of energies of atomic Rydberg interactions. Singh et al. \cite{singh2022dual} demonstrated experimental feasibility to create a dual-element atomic array with individual control of single $^{87}$Rb and $^{133}$Cs atoms which was verified for 512 trapping sites with negligible crosstalk for $10$~\si{\mu m} ($\simeq 7$~\si{\mu m}) distance between  positions of atoms of the same (or different) species. This heteronuclear architecture allowed Rb and Cs atoms to be trapped, cooled, loaded to the optical dipole trap and controlled independently.

Dual-element array of homonuclear architecture \cite{sheng2022defect,bluvstein2022quantum}, heteronuclear atomic architecture \cite{singh2022dual}, or an array of polar molecules interacting with Rydberg atoms \cite{zhang2022quantum}, overcome the challenge to all architectures of quantum systems by allowing large system sizes and low crosstalk. Theoretical calculations of interspecies couplings in \cite{walker2008consequences, beterov2015rydberg} anticipated obtaining high fidelity quantum non-demolition state measurements with low cross-talk in qubit arrays. Heuristic connectivity optimization algorithms were proposed to provide the fewest number of atom moves to rearrange the loaded dual-species atoms arrays \cite{tao2022efficient} or to arrange atomic arrays with user-defined geometries \cite{sheng2022defect}. Dual-element trapping can simulate models of complex structure in quantum many-body physics, leading to advantageous encoding of the maximum independent set problem using Rydberg atom arrays \cite{ebadi2022quantum, nguyen2022quantum, byun2022finding, kim2022rydberg}. This architecture can be promising for achievement of the quantum supremacy~\cite{arute2019quantum}).

Our work  focuses on calculating the obtained fidelity for the CNOT gates  which are implemented parallelly in a large atomic arrays. We considered sequential creation of entanglement between all neighboring qubits which is necessary for surface codes, creation of cluster states and other applications. We used the scheme of two-qubit gates based on electromagnetically induced transparency (EIT), proposed by M\"{u}ller et al.~\cite{muller2009mesoscopic}. This scheme was experimentally implemented in Ref.~\cite{mcdonnell2022demonstration}. The coherent transport of atoms was demonstrated in Ref.~\cite{bluvstein2022quantum}. The dual element architecture of the atomic arrays~\cite{singh2022dual} is related to possible experimental realization of our proposal on heteronuclear parallel implementation of CNOT gates~\cite{farouk2022parallel}.

The paper is organized as follows: in section~\ref{section_physics}, a scheme of a CNOT gate based on EIT is described, We show the structure of the atomic energy levels and the sequence of the laser pulses. In section~\ref{section-trans-fidelity}, the gate fidelity is studied depending on the average speed of transporting the array of ancillas from one position to the other and on the minimal distance between control and target atoms. In section~\ref{section-mutual-information}, the mutual information and Renyi entropy are discussed. In section~\ref{section-conclusion}, the results of the manuscript are concluded.

\section{Physical model \label{section_physics}}

\begin{figure}[htb!]\centering
		\subfloat[]{
			\includegraphics[width=3.5cm,height=3.5cm]{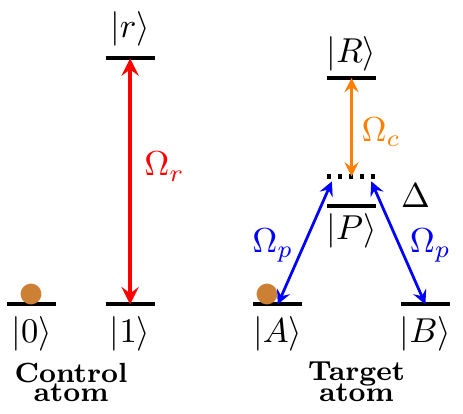}
		}
		\subfloat[]{
			\includegraphics[width=3.5cm,height=3.5cm]{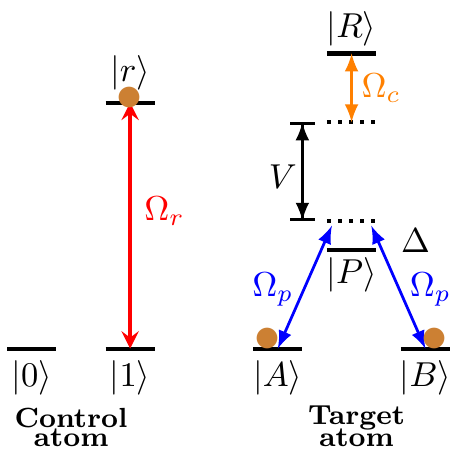}
		}
		\\
		\subfloat[]{
			\includegraphics[width=4.5cm,height=3.5cm]{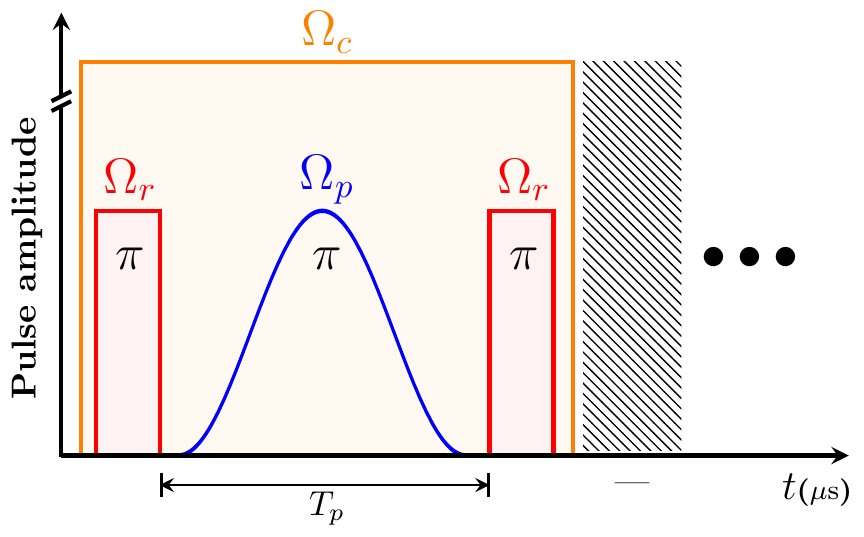}
		}
		\subfloat[]{
			\includegraphics[width=3.5cm,height=3.5cm]{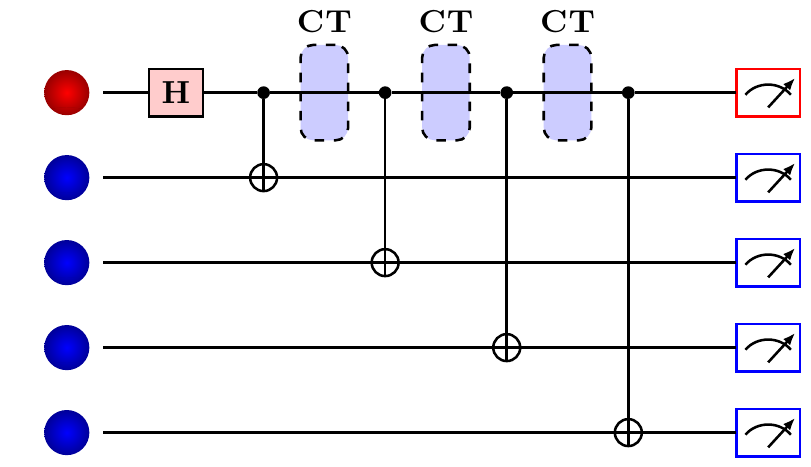}
		}\\
		\subfloat[]{
			\includegraphics[width=2cm,height=2cm]{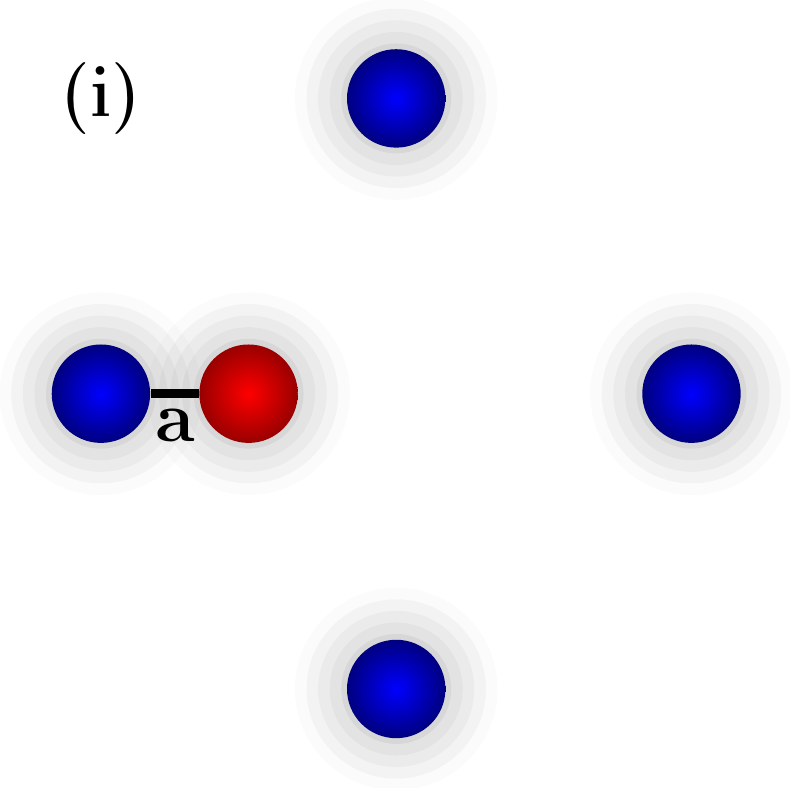}
			
			\includegraphics[width=2cm,height=2cm]{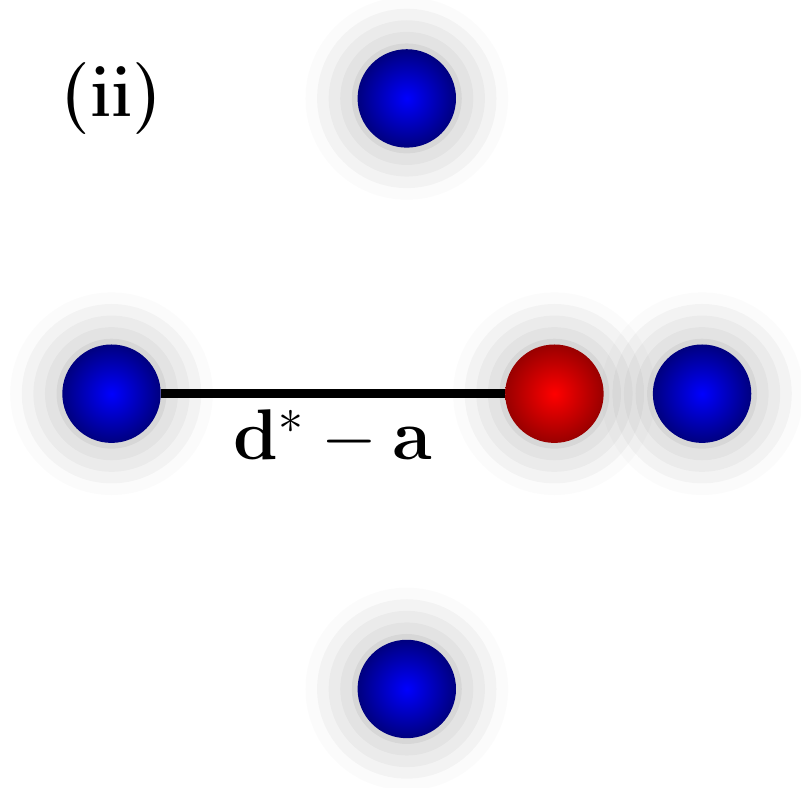}
			
			\includegraphics[width=2cm,height=2cm]{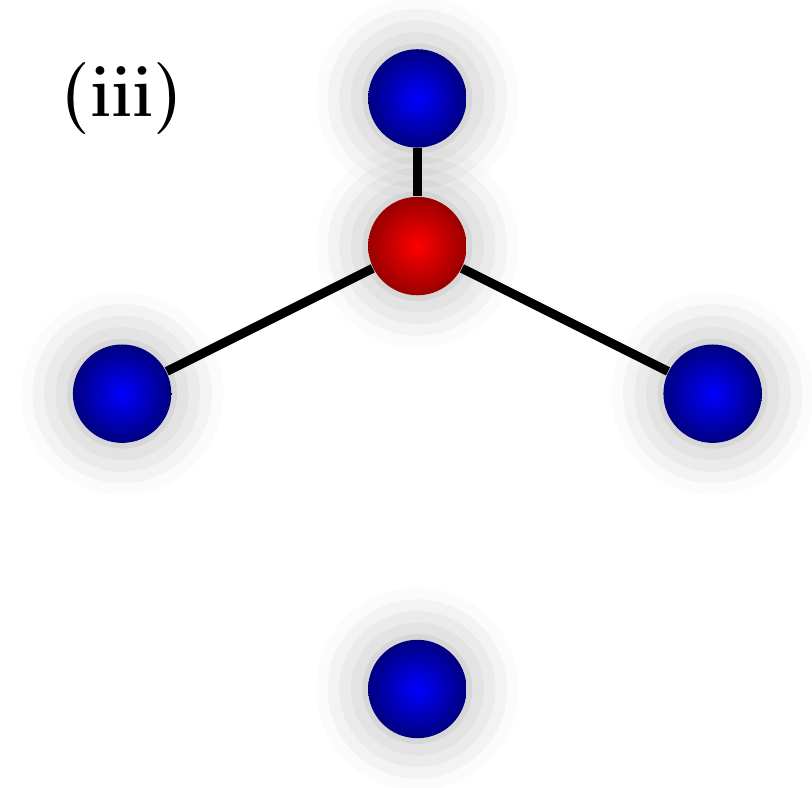}
			
			\includegraphics[width=2cm,height=2cm]{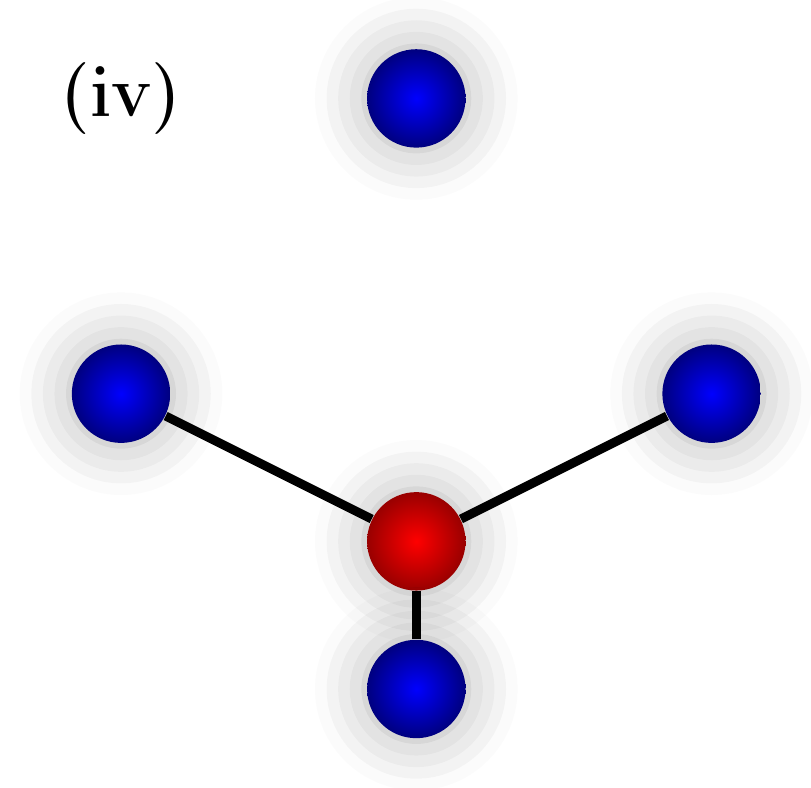}
		}
		\caption{Scalable architecture of neutral heteronuclear atoms to perform CNOT gate using EIT protocol by coherently transporting the control atom (ancilla qubit) between four target atoms (data qubits). 
		\textbf{(a)} Due to the effect of EIT, the population transfer between $|A\rangle$ and $|B\rangle$ can be blocked efficiently. 
		\textbf{(b)} The control atom is excited to Rydberg state $|r\rangle$, allowing for the interaction between the Rydberg states which violates the EIT condition and allows the population transfer between the ground states of target atoms. 
		\textbf{(c)} The sequence of pulses applied to control and target atom $\mathcal{N}$. The gray-shaded region is a time gap during which coherent transport occurs.
		\textbf{(d)} The circuit representation of the implementation of CNOT gates while coherently transporting (CT) the control atom (ancilla qubit) among target atoms (data qubits).
		\textbf{(e)} The transportation scheme of the control atom between the target atoms. The minimum distance between the control atom and any of the target atoms is $a> R_{\tiny\textrm{LR}}$, the distance between the nearest target atoms and is equal to $d = R_{\tiny\textrm{TT}} = 60$~\si{\mu m}.
		} \label{Fig1-spatialconfiguration}
\end{figure}

\begin{figure*}[htb!]\centering
	\subfloat[]{
		\includegraphics[width=5.5cm,height=4.5cm]{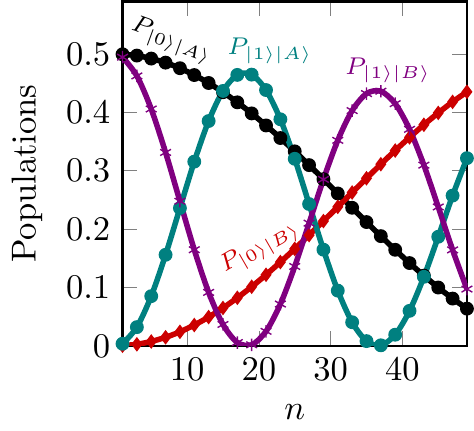}
	}
	\subfloat[]{
		\includegraphics[width=5.5cm,height=4.5cm]{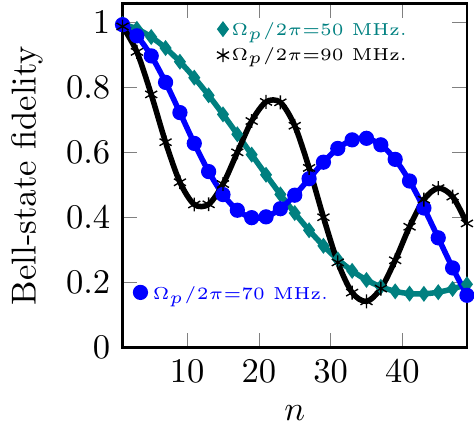}
	}
	\subfloat[]{
		\includegraphics[width=5.5cm,height=4.5cm]{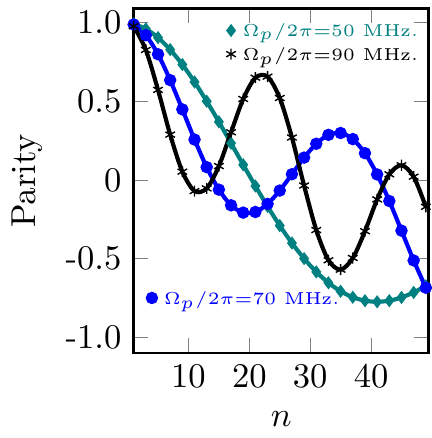}
	}
	\caption{\textbf{(a)} The populations of the logical computational states for only one target atom as a function of the number of cycles $n \in (2\mathbb{N}-1)$ considering the system initially prepared in the superposition of block and transfer gates $\frac{1}{\sqrt{2}}\left( |0\rangle+ |1\rangle \right) |A\rangle $, $\Omega_{p}=2\pi\times70$~\si{MHz}.
		\textbf{(b-c)} Bell-state fidelity and parity oscillation $\mathcal{P}=2 (P_{\text{$|$0$\rangle$$|$A$\rangle$}} + P_{\text{$|$1$\rangle$$|$B$\rangle$}} ) - 1$, as a function of the number of cycles as in (c). Parity has a calculated amplitude of $98.86\%$, $97.77\%$ for CNOT implementation with only one cycle i.e. $n=1$ considering $\Omega_{p}=2\pi \times 70$~\si{MHz} and $2\pi \times 90$~\si{MHz}, respectively. $a=5$~\si{\mu m}, $\Omega_{c}=2.5~\Omega_{p}$, and gate time $\tau_{n}=n \, (2\,T_r+T_p)+ (n-1)\,T_{gap}$.} 
	\label{Fig2-OneAtomMultiPulses}
\end{figure*}

We consider a heteronuclear atomic architecture by using an array of cesium  $^{133}$Cs atoms in a trap generated by AOD. These atoms act as control atoms (ancillas) with ground states $|0\rangle=|6S_{1/2},F=3, m_F=0\rangle$, and $|1\rangle=|6S_{1/2},F=4, m_F=0\rangle$ which are coupled to Rydberg state $|r\rangle=|81S_{1/2},m_j=-1/2\rangle$ by a sharp  $\pi$-pulse with Rabi frequency $\Omega_{r}(t)$ applied for $T_{r}$~(\si{\mu s}) [see Fig.~\ref{Fig1-spatialconfiguration}(a,b)]. The pulse shape is defined as

\begin{equation}\label{OmegaRfunction}
	\Omega_{r}^{(\mathcal{N})}(t)=\left\{
	\begin{array}{ll}
		0,				& t<\mathds{T}_{0,0}^{(\mathcal{N})}.\\
		\dfrac{\pi}{T_{r}},		 &	\mathds{T}_{0,0}^{(\mathcal{N})} \leq t\leq \mathds{T}_{1,0}^{(\mathcal{N})}.\\
		0,				& \mathds{T}_{1,0}^{(\mathcal{N})} < t < \mathds{T}_{1,1}^{(\mathcal{N})}.\\
		\dfrac{\pi}{T_{r}},		 &	\mathds{T}_{1,1}^{(\mathcal{N})} \leq t \leq \mathds{T}_{2,1}^{(\mathcal{N})}.\\
		0,				& t > \mathds{T}_{2,1}^{(\mathcal{N})}.
	\end{array}
	\right.
\end{equation}	
where $$\mathds{T}_{j,k}^{(\mathcal{N})}=\mathcal{T}_{\mathcal{N}}+j \, T_{r}+ k \, T_{p},$$ 
and
$$\mathcal{T}_{\mathcal{N}}=\sum_{b}^{\mathcal{N}}(b-1)\big[ 2 \, T_{r} + T_{p} + T_{gap} \big].$$
$T_{gap}$ , and $T_{p}$ are the transportation time, and the operating time of a Raman laser pulse, $\mathcal{N}$ is the index denoting the target atom for which the CNOT gate is performed, the indices $j = 0,1,2$ and $k = 0,1$ are used to indicate the on and off times of laser pulses. The Rydberg state of control atom has a lifetime of $\tau_r=548$~\si{\mu s}. The Hamiltonian of the control atom reads 
\begin{equation}
	\hat{H}_{\text{\tiny C}}=\frac{1}{2}\hbar\,\bigg[\Omega_{r}^{(\mathcal{N})}(t) \left( |1\rangle \langle r|+h.c. \right) -i \, \gamma_{r} |r\rangle\langle r|  \bigg],
\end{equation}
Here $\gamma_{r}=1/\tau_r$ is the rate of spontaneous decay of the Rydberg state.An array of rubidium $^{87}$Rb target atoms is trapped on the same plane as an array of ancillas where the traps generated by SLM. For  target (data) atoms the logical ground states are $|A\rangle = |5S_{1/2},F=1, m_{F}=0\rangle$, and $|B\rangle = |5S_{1/2},F=2, m_{F}=0\rangle$. The ground states of a target atom are coupled to the intermediate state $|P\rangle=|6P_{3/2}, m_j=3/2\rangle$ by a smooth Raman $\pi$-pulse  with Rabi frequency $\Omega_{p}(t)$ and the profile  
$$\Omega_{p}(t)= \sqrt{\frac{16\pi\Delta}{3T_{p}}} \sin^{2}\left(\frac{\pi \, t}{T_{p}}\right).$$

This pulse acts only on target atoms and is described as 
$$\frac{1}{2 \, \Delta}\int_{0}^{T_{p}}\Omega_{p}^{2}(t) dt = \pi, $$ where $\Delta$ is the detuning  from the resonance between the ground states and intermediate excited state. To highly suppress the effect of the spontaneous decay of the intermediate state, Raman detuning $\Delta$ has to be much larger than the inverse of the decay rate $\gamma_{p}$ \cite{mansell2014cold}. The intermediate state $|P\rangle$ has a lifetime $\tau_p=26.4$~\si{ns} and is coupled to Rydberg state $|R\rangle=|77S_{1/2}, m_j=1/2\rangle$ by radiation with the Rabi frequency $\Omega_{c}(t)$ [see Fig.~\ref{Fig1-spatialconfiguration}(a,b)]. Explicit forms of the time-dependent laser pulses of Raman $\Omega_{p}(t)$ and Rabi $\Omega_{c}(t)$ functions, are defined as follows [see Fig.~\ref{Fig1-spatialconfiguration}(c)]
\begin{equation}\label{OmegaPfunction}
	\Omega_{p}^{(\mathcal{N})}(t)=
	\left\{
	\begin{array}{ll}
		0,	& t<\mathds{T}_{1,0}^{(\mathcal{N})}.\\
		\sqrt{\frac{16\pi\Delta}{3 \, T_{p}}}\sin^2 (\dfrac{\pi \, \bar{t}}{T_{p}}),& \mathds{T}_{1,0}^{(\mathcal{N})} \leq t \leq \mathds{T}_{1,1}^{(\mathcal{N})}.\\
		0,	& t > \mathds{T}_{1,1}^{(\mathcal{N})}.
	\end{array}
	\right.
\end{equation}
where $\bar{t}=t-\mathds{T}_{1,0}^{(\mathcal{N})} $, and 
\begin{equation}\label{OmegaCfunction}
	\Omega_{c}^{(\mathcal{N})}(t)=\left\{
	\begin{array}{ll}
		0,				& t<\mathds{T}_{0,0}^{(\mathcal{N})}.\\
		\Omega_{c},		 &	\mathds{T}_{0,0}^{(\mathcal{N})} \leq t\leq  \mathds{T}_{2,1}^{(\mathcal{N})}.\\
		0,				& t> \mathds{T}_{2,1}^{(\mathcal{N})}.
	\end{array}
	\right.
\end{equation}
The Hamiltonian of a target atom $j$ reads 
\begin{equation}
	\begin{split}
		\hat{H}_{\text{\tiny T}_j}=&\frac{1}{2}\hbar\, \bigg[ \Omega_{p}^{(\mathcal{N})}(t) \bigg( | A \rangle\langle P | + | B \rangle\langle P | + h.c. \bigg)  \\& +\Omega_{c}^{(\mathcal{N})}(t) \bigg( | P \rangle \langle R | + h.c. \bigg)-\,\bigg( 2 \Delta + i \gamma_{p} \bigg)\otimes \\& \hskip 5cm \otimes | P \rangle\langle P |\bigg],
	\end{split}
\end{equation} 
where $\gamma_{p}=1/\tau_p$ is the rate of spontaneous decay of the intermediate $|P\rangle$ state of the target atom.

\begin{figure*}[t]\centering 
	\subfloat[$\mathcal{N}=1$]{
		\includegraphics[width=0.23\textwidth, height=4cm]{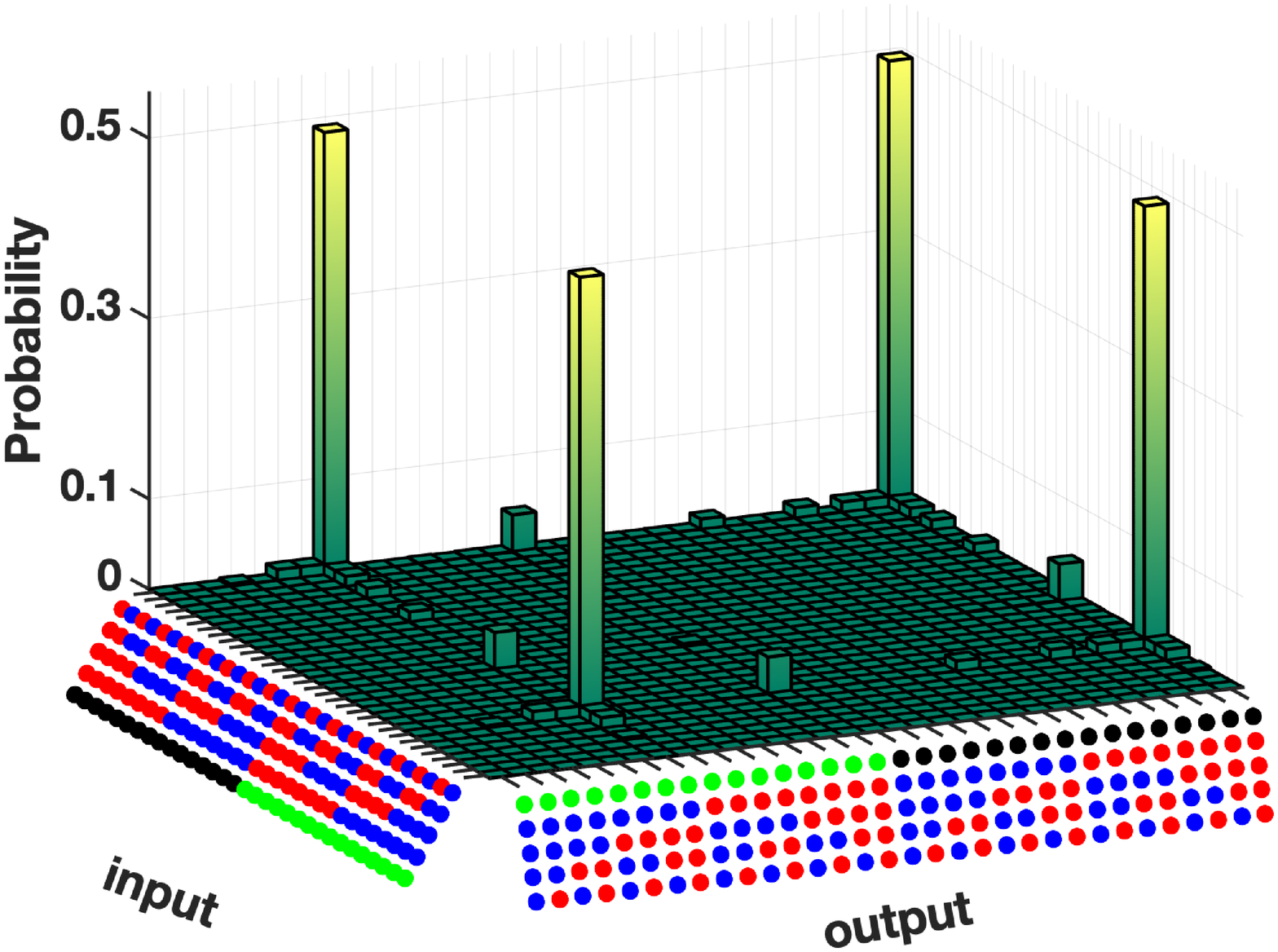}
		\begin{picture}(0,0)
			\put(-48,85){\tikz\draw[color=white, fill=red] (2.1,2.8) circle [radius=0.1] node {\tiny 1};}
			\put(-27,73){\tikz\draw[color=white, fill=red] (2.1,2.8) circle [radius=0.1] node {\tiny 2};}
			\put(-74,68){\tikz\draw[color=white, fill=red] (2.1,2.8) circle [radius=0.1] node {\tiny 3};}
			\put(-95,80){\tikz\draw[color=white, fill=red] (2.1,2.8) circle [radius=0.1] node {\tiny 4};}
		\end{picture}
	}
	\subfloat[$\mathcal{N}=2$]{
		\includegraphics[width=0.23\textwidth, height=4cm]{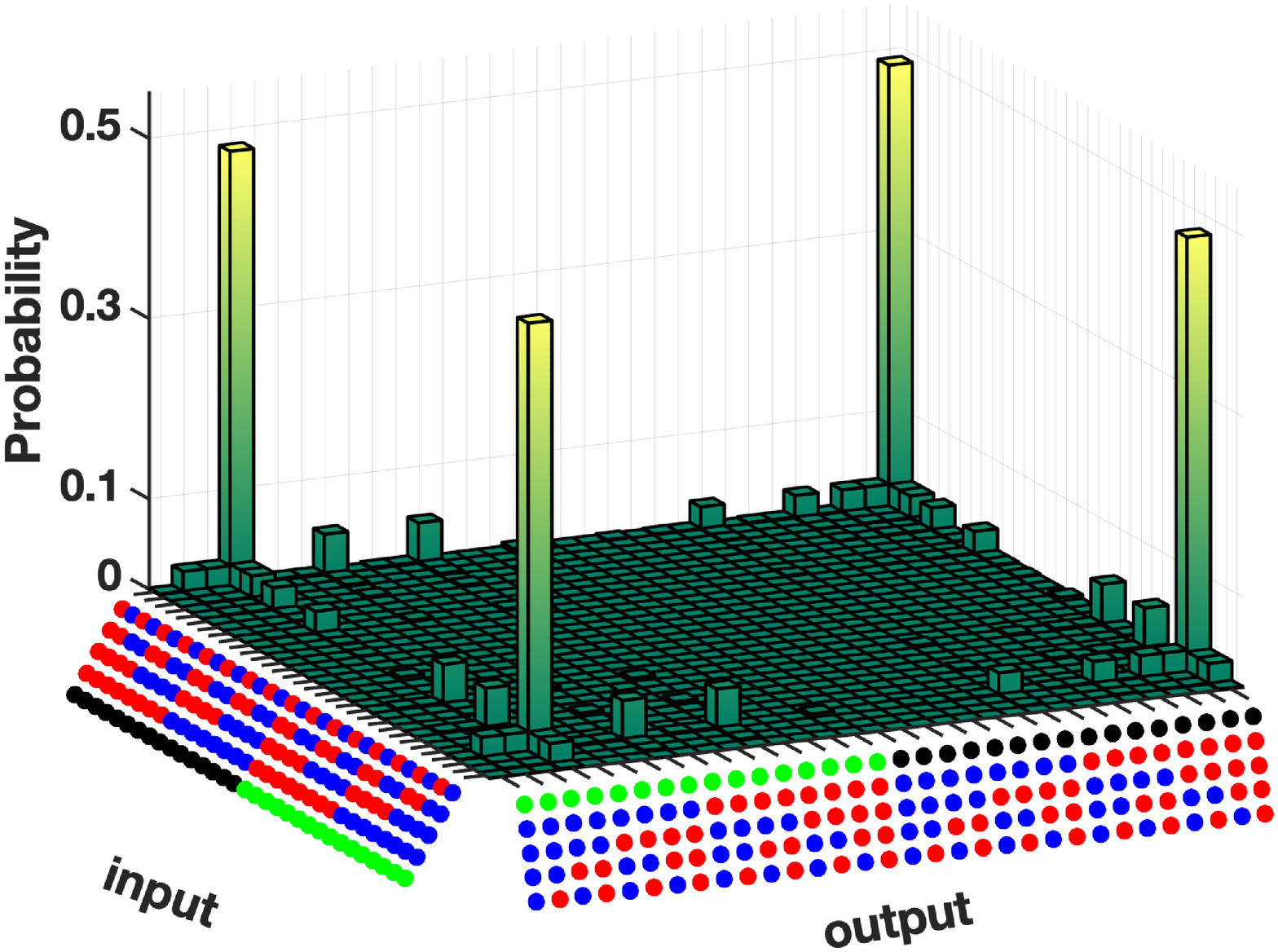}
	}
	\subfloat[$\mathcal{N}=3$]{
		\includegraphics[width=0.23\textwidth, height=4cm]{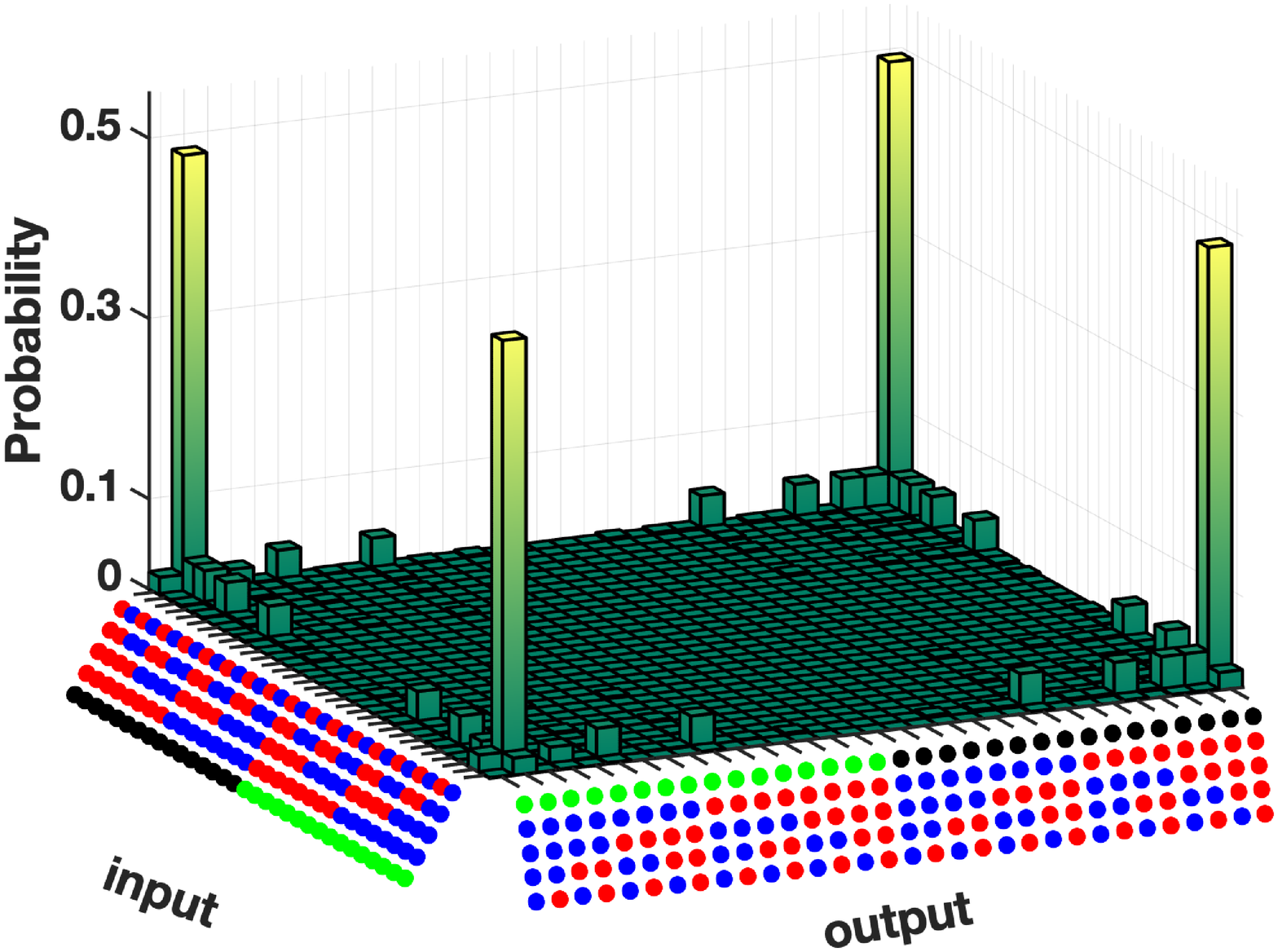}
	}
	\subfloat[$\mathcal{N}=4$]{
		\includegraphics[width=0.23\textwidth, height=4cm]{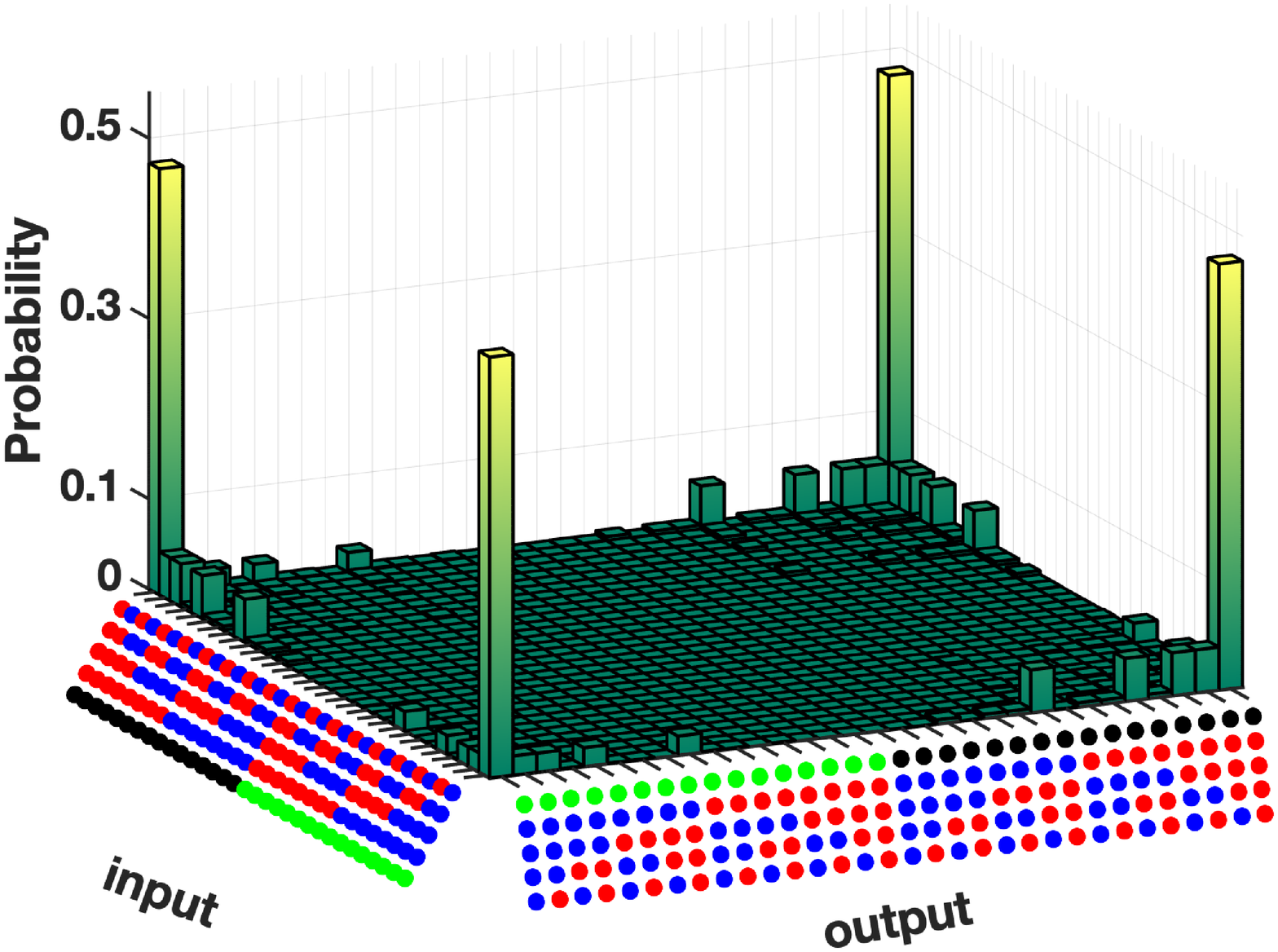}
	}\\
	\subfloat[$\mathcal{N}=1$]{
		\includegraphics[width=0.23\textwidth, height=4cm]{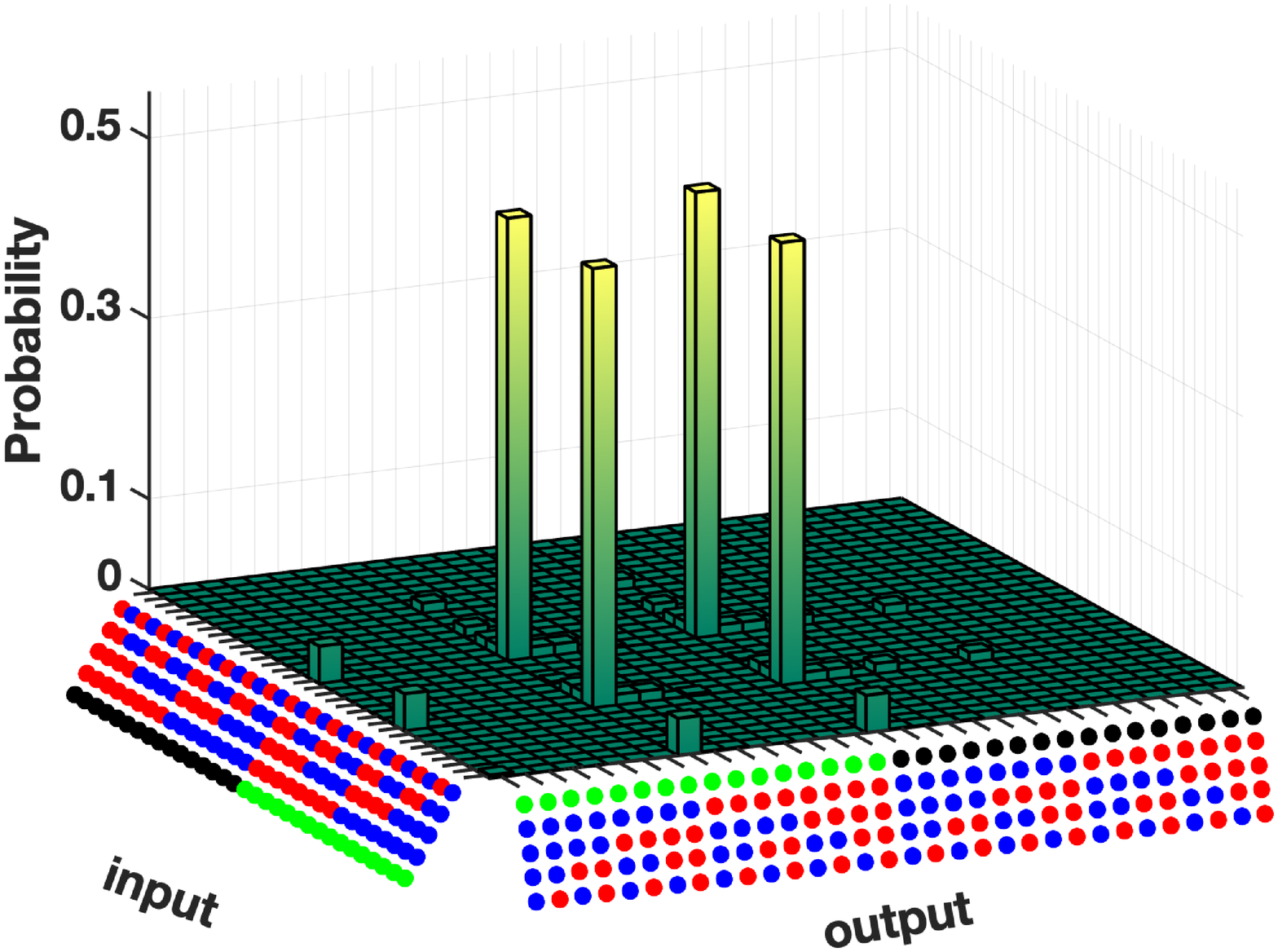}
		\begin{picture}(0,0)
			\put(-64,75){\tikz\draw[color=white, fill=red] (2.1,2.8) circle [radius=0.1] node {\tiny 1};}
			\put(-55,71){\tikz\draw[color=white, fill=red] (2.1,2.8) circle [radius=0.1] node {\tiny 2};}
			\put(-73,68){\tikz\draw[color=white, fill=red] (2.1,2.8) circle [radius=0.1] node {\tiny 3};}
			\put(-80,73){\tikz\draw[color=white, fill=red] (2.1,2.8) circle [radius=0.1] node {\tiny 4};}
		\end{picture}
	}
	\subfloat[$\mathcal{N}=2$]{
		\includegraphics[width=0.23\textwidth, height=4cm]{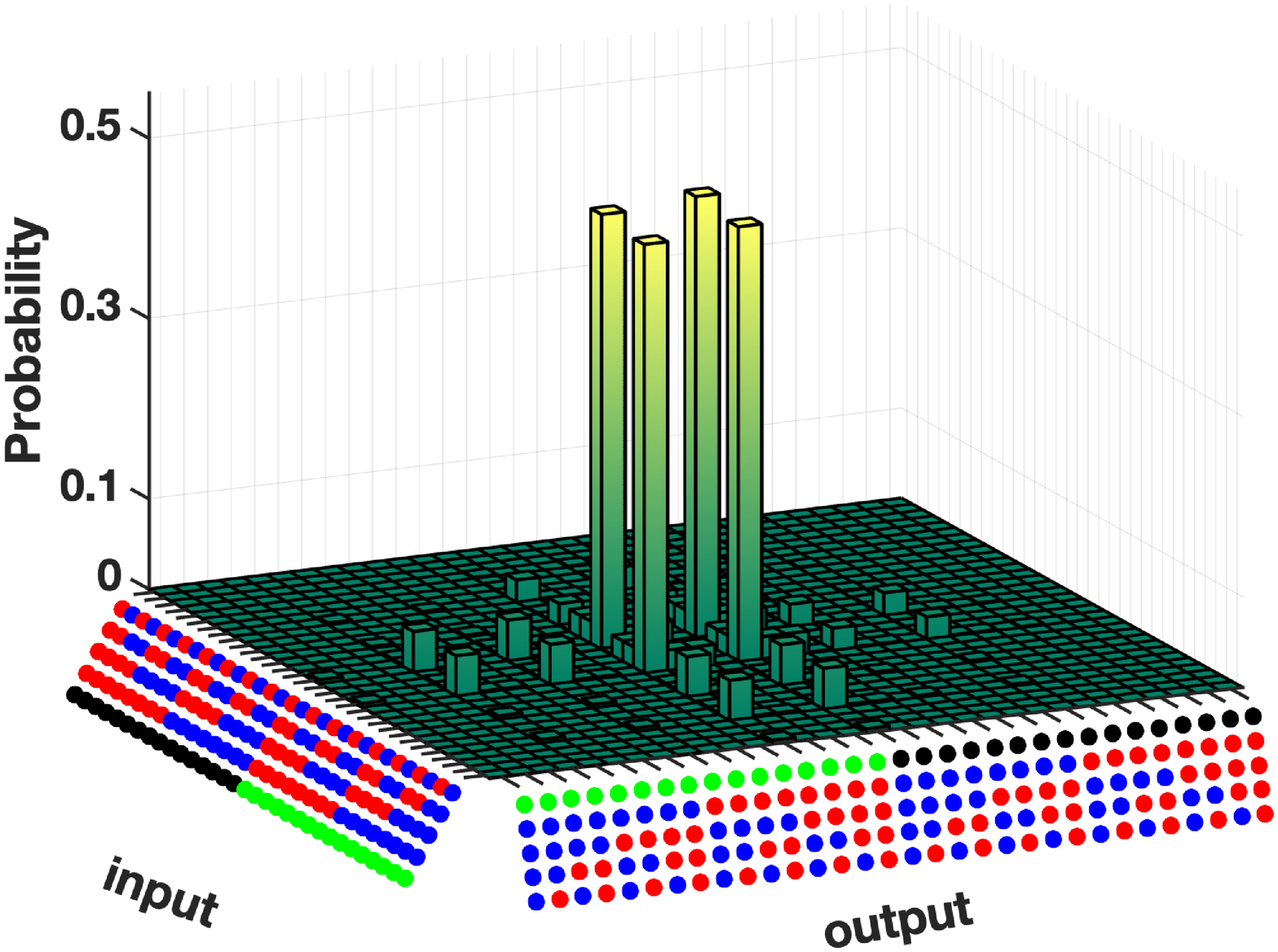}
	}
	\subfloat[$\mathcal{N}=3$]{
		\includegraphics[width=0.23\textwidth, height=4cm]{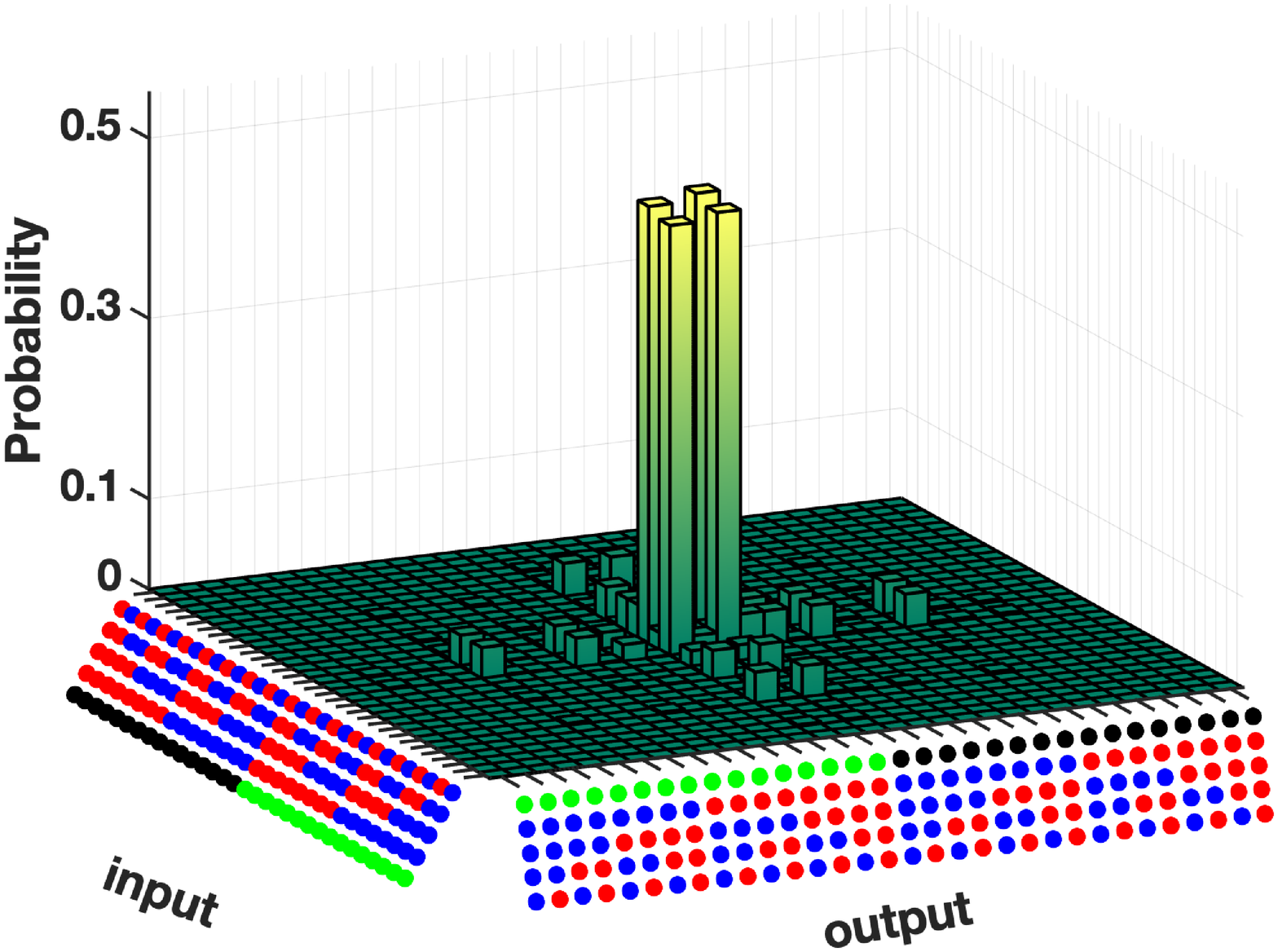}
	}
	\subfloat[$\mathcal{N}=4$]{
		\includegraphics[width=0.23\textwidth, height=4cm]{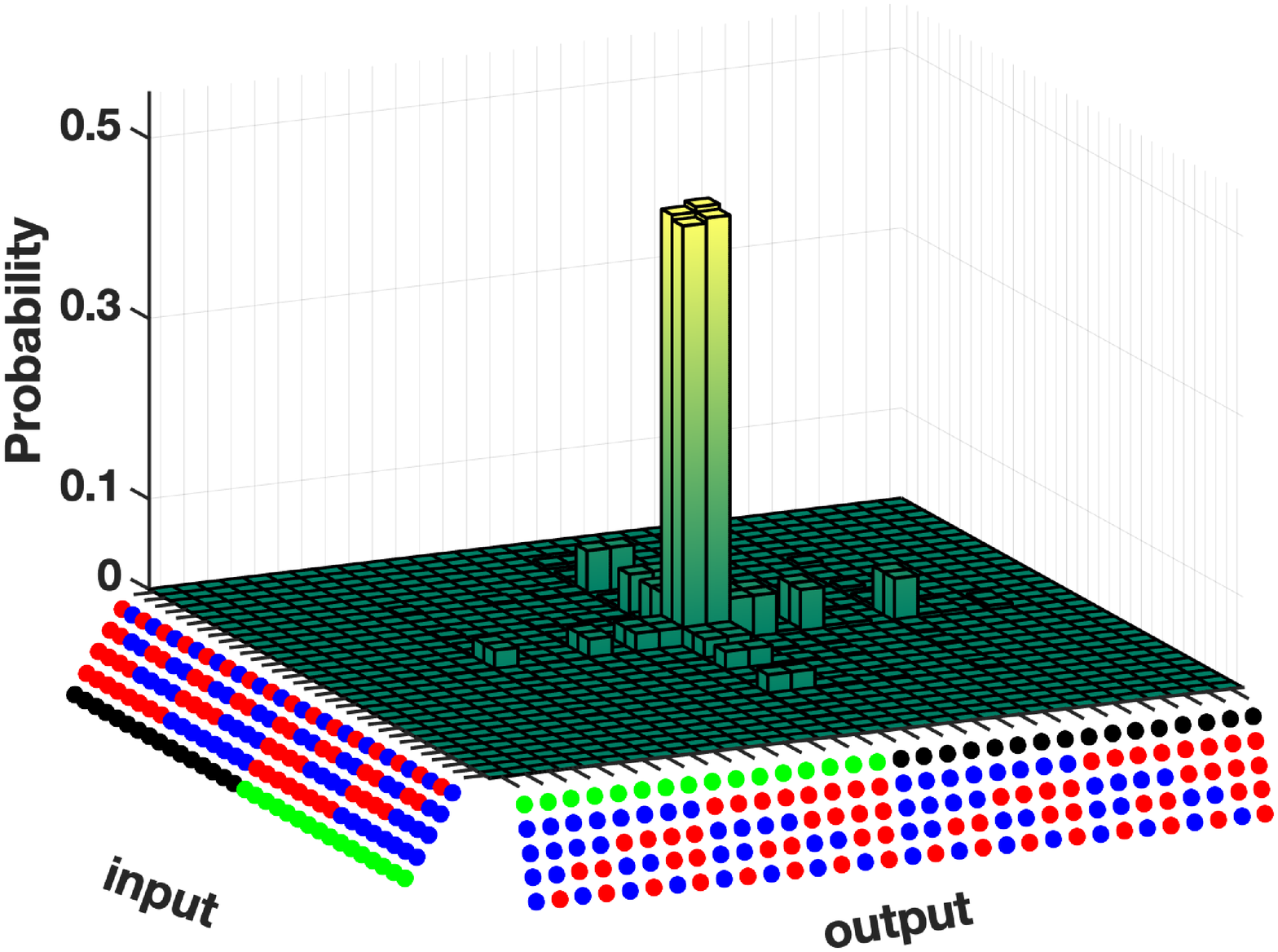}
	}
	\caption{Truth tables of the generation of GHZ-state after implementing CNOT$^{\mathcal{N}}$ gates of each of target atoms. (a-d) $|\psi_{0}\rangle=\frac{1}{\sqrt{2}}\left( |0\rangle  + |1\rangle \right) | AAAA\rangle \rightarrow \frac{1}{\sqrt{2}}\left( |0\rangle | AAAA\rangle + |1\rangle | B \rangle^{\otimes \mathcal{N}} | A \rangle^{\otimes (N-\mathcal{N})} \right)$, (e-h) $|\psi_{0}\rangle=\frac{1}{\sqrt{2}}\left( |0\rangle+ |1\rangle \right)  | BBBB\rangle \rightarrow \frac{1}{\sqrt{2}}\left( |0\rangle | BBBB\rangle+ |1\rangle | A \rangle^{\otimes \mathcal{N}} | B \rangle^{\otimes (N-\mathcal{N})} \right)$. Ground states $|0\rangle$, $|1\rangle$, $|A\rangle$, and $|B\rangle$ are notated as \textcolor{black}{\Large $\bullet$}, \textcolor{green}{\Large $\bullet$}, \textcolor{red}{\Large $\bullet$}, and \textcolor{blue}{\Large $\bullet$}, respectively. Total number of target atoms $N=4$, $T_{\text{\tiny gap}}=1.09$~\si{\mu s}, $T_{\text{\tiny r}}=16.6$~\si{n s}, $a=5$~\si{\mu m}, $d=60$~\si{\mu m}, $\Delta=2\pi\times1200$~\si{MHz}, $\Omega_{p}=2\pi\times70$~\si{MHz}, and $\Omega_{c}=2.5\, \Omega_{p}=2\pi\times175$~\si{MHz}.}
	\label{Fig3:truthtable}
\end{figure*}

For the given definitions of laser pulses $\Omega_{r}(t), \Omega_{p}(t)$, and $\Omega_{c}(t)$, if the control atom is prepared in the ground state $|0\rangle$, there will be no Rydberg excitation of the control atom and therefore no interaction between the control and target atoms. For proper choice of the value  $\Omega_{c}$, the population transfer between ground states $|A\rangle$ and $|B\rangle$ of target atom $j$ will be blocked [see Fig.~\ref{Fig1-spatialconfiguration}(a)]. If the control atom is prepared in the ground state $|1\rangle$, the laser pulse $\Omega_{r}(t)$ will excite the atom into Rydberg state  $|r\rangle$ and the Rydberg interaction of control and target atoms $j$ will  result in a shift of the energy of Rydberg state $|R\rangle$ by $V_{\text{\tiny CT}_j}>0$, This lifts the condition for two-photon resonance in the target atom and allows the transfer of population between its ground states $|A\rangle$ and $|B\rangle$. This population transfer between ground states of target atom can be regarded as a conditional CNOT gate [see Fig.~\ref{Fig1-spatialconfiguration}(b)]. This model of CNOT was proposed by M\"{u}ller et al. \cite{muller2009mesoscopic}. The interaction Hamiltonian between control and target atom $j$ reads $$\hat{H}_{\text{\tiny CT}_j}=V_{\text{\tiny CT}_j} |r\rangle \langle r| \otimes | R \rangle_j \langle R|,$$
where $V_{\text{\tiny CT}_j}=\frac{C_{3}}{R^{3}}$, $C_3/2\pi=14.25$~\si{GHz}~\si{\mu m^3} and $R_{\text{\tiny LR}}=1.9$~\si{\mu m}\footnote{In our simulation, interatomic distances between control and target atoms $R_{\text{CT}_j}$ are time-dependent functions determining the actual position of the control atom during the implementation process.}. All values of interaction energies or lifetimes are calculated using Alkali Rydberg Calculator (ARC) Python package  \cite{vsibalic2017arc} and the detailed calculations were reported in our paper \cite{farouk2022parallel}.

The configuration of trapped atoms in both traps is designed in such a way that the distance between atoms of the same atomic species is large enough to suppress any homonuclear Rydberg interactions within the array $$\hat{H}_{\text{\tiny TT}}=\sum_{j \neq k}^{N} V_{\text{\tiny T}_j\text{\tiny T}_k} |R\rangle_j \langle R | \otimes |R\rangle_k \langle R |,$$ and to suppress heteronuclear Rydberg interaction between the control and distant target atoms $k$. Our gate sequence requires $a=R_{\text{\tiny CT}_j} << R_{\text{\tiny CT}_k}$ ($j\neq k$). The interaction between rubidium target atoms lies in the regime of van der Waals (vdW), i.e. $$V_{\text{\tiny T}_j\text{\tiny T}_k}=\frac{C_{6}}{R^{6}},$$
with $C_{6}/2\pi=2036$~\si{GHz}~\si{\mu m}$^{6}$ and $R_{\text{\tiny LR}}=1.8$~\si{\mu m}\footnote{$R_{\text{T}_j \text{T}_k}$ has a constant value, since the target atoms are trapped in a static trap.}. Considering the distance between two neigbouring target atoms $d>20$ \si{\mu m}, then $V_{\text{\tiny T}_j\text{\tiny T}_k} <<V_{\text{\tiny CT}_j} $. Interaction between heteronuclear atoms is dipole-dipole (d-d) and not van der Waals for interatomic distance between control and the nearest target atom $R_{\text{\tiny CT}_j}< R_{\text{\tiny vdW}}=31$~\si{\mu m}. The gate time $$\tau_{N}=N \, (2 \, T_{r}+T_{p}) + (N-1) \,T_{\text{\tiny gap}},$$ where $N=4$ is the number of target atoms. 

In Fig.~\ref{Fig2-OneAtomMultiPulses}, we study the simplest case when there is only one target atom [see Fig.~\ref{Fig1-spatialconfiguration}(d)]. The control atom after applying an Hadamard gate will be prepared in a superposition of ground states $$H \, |0\rangle=\frac{1}{\sqrt{2}}(|0\rangle+|1\rangle).$$ 
The sequence of applied pulses described above should generate Bell states. Hadamard gate can be realized on Bloch sphere representation by performing a rotation of $90^{\circ}$ over $Y$-axis followed by a rotation of $180^{\circ}$ over $X$-axis. It is noted that the population of the transfer gate is being swapped $|1\rangle|A\rangle \leftrightarrow |1\rangle|B\rangle$) regularly and the maximum value is decreasing compared to the initial state. The rate of populations swapping depends on the value of $\Omega_{p}$ as shown in Fig.~\ref{Fig2-OneAtomMultiPulses}(b).

 \begin{figure}[htb!]\centering
	\includegraphics[width=\columnwidth, height=6cm]{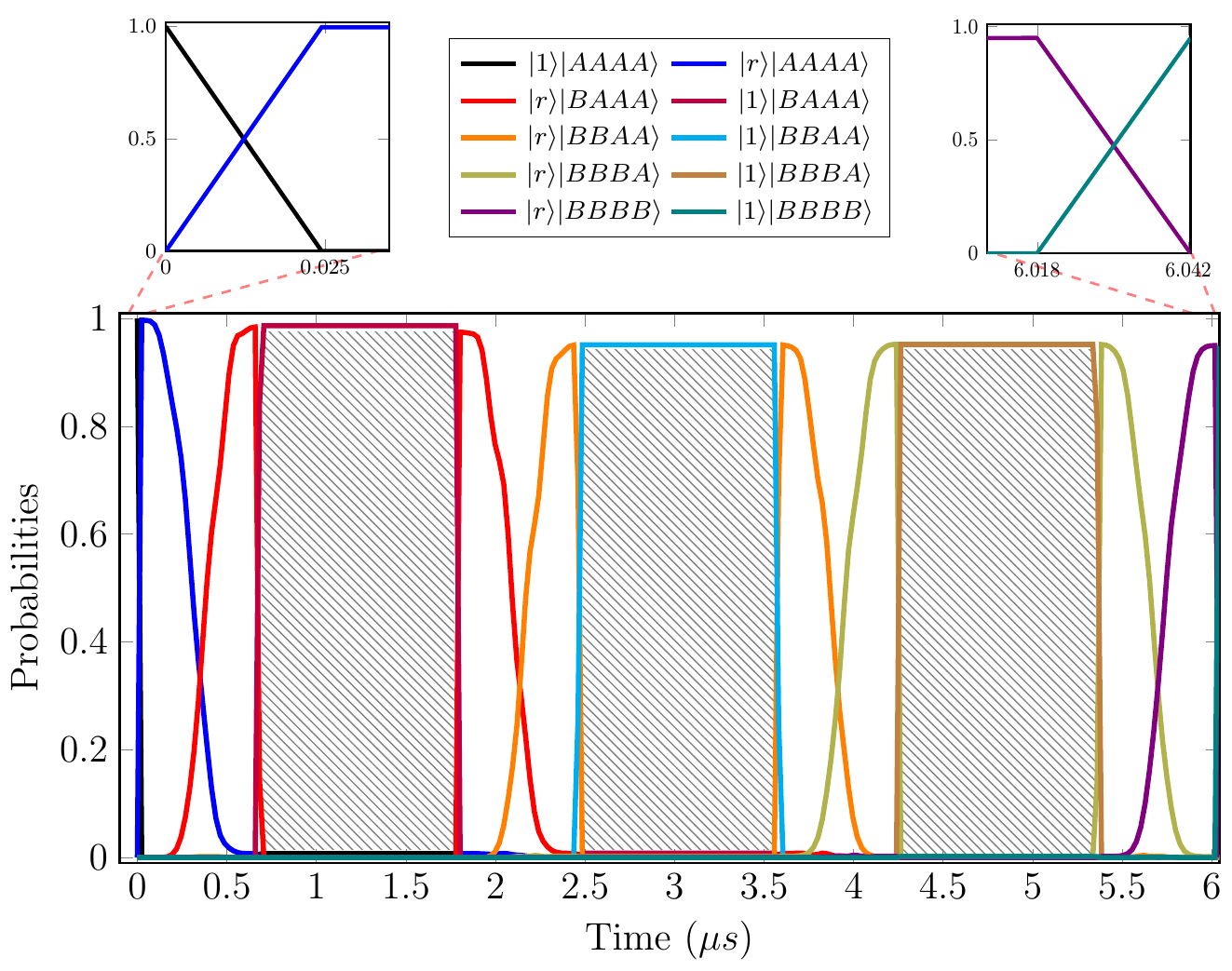}
	\caption{The population transfer from $|1\rangle |AAAA\rangle$ to $|1\rangle |BBBB\rangle$, using the same initial values as in Fig.~\ref{Fig3:truthtable}. The grey-shaded parts represent the gap time $T_{gap}$ (see main text).}
	\label{Fig4:CT-populations}
\end{figure}

In Fig.~\ref{Fig3:truthtable}, we show the truth-table for the generation of GHZ-state 
\begin{equation}
	\begin{split}
		|\psi_{0}\rangle=&\frac{1}{\sqrt{2}}\left( |0\rangle  + |1\rangle \right) | AAAA\rangle \rightarrow \\& \frac{1}{\sqrt{2}}\left( |0\rangle | AAAA\rangle + |1\rangle | B \rangle^{\otimes \mathcal{N}} | A \rangle^{\otimes (N-\mathcal{N})} \right),
	\end{split}
\end{equation}
for the upper panel, and
\begin{equation}
	\begin{split}
		|\psi_{0}\rangle=&\frac{1}{\sqrt{2}}\left( |0\rangle+ |1\rangle \right)  | BBBB\rangle \rightarrow \\& \frac{1}{\sqrt{2}}\left( |0\rangle | BBBB\rangle+ |1\rangle | A \rangle^{\otimes \mathcal{N}} | B \rangle^{\otimes (N-\mathcal{N})} \right)
	\end{split}
\end{equation}
for the lower panel, considering array of the control atoms is being transported as in Fig.~\ref{Fig1-spatialconfiguration}(e). Column 1 (indicated by red circle) in the calculated truth tables in Fig.~\ref{Fig3:truthtable}(a) corresponds to the preservation of the system in its original state $|0\rangle|AAAA\rangle\xrightarrow{\text{CNOT}} |0\rangle|AAAA\rangle$. It can be seen that its probability amplitude practically does not decrease under the action of multiple laser pulses and coherent transport, shown in Fig.~\ref{Fig2-OneAtomMultiPulses}(a). We tracked the population transfer from $ | 1 \rangle | AAAA \rangle $ to $ | 1 \rangle | BBBB \rangle$, as depicted in Fig.~\ref{Fig4:CT-populations}, through the route

$|1\rangle|AAAA\rangle$ $\xrightarrow{\Omega_{r}(t)}$ $|r\rangle|AAAA\rangle $ $ \xrightarrow{\text{SWAP}}$ $|r\rangle|BAAA\rangle $ $\xrightarrow{\Omega_{r}(t)}$ $|1\rangle|BAAA\rangle $ $\xrightarrow{\text{CT}} $ $ |1\rangle|BAAA\rangle $ $\xrightarrow{\Omega_{r}(t)}$ $|r\rangle|BAAA\rangle $ $\xrightarrow{\text{SWAP}}$ $|r\rangle|BBAA\rangle $ $\xrightarrow{\Omega_{r}(t)}$ $|1\rangle|BBAA\rangle $ $\xrightarrow{\text{CT}}$ $|1\rangle|BBAA\rangle $ $\xrightarrow{\Omega_{r}(t)}$ $|r\rangle|BBAA\rangle $ $\xrightarrow{\text{SWAP}}$ $|r\rangle|BBBA\rangle $ $\xrightarrow{\Omega_{r}(t)}$ $|1\rangle|BBBA\rangle $ $\xrightarrow{\text{CT}}$ $|1\rangle|BBBA\rangle $ $\xrightarrow{\Omega_{r}(t)}$ $|r\rangle|BBBA\rangle $ $\xrightarrow{\text{SWAP}}$ $|r\rangle|BBBB\rangle $ $\xrightarrow{\Omega_{r}(t)}$ $|1\rangle|BBBB\rangle$ corresponds to column 3 (indicated by red circle) in Fig.~\ref{Fig3:truthtable}.


\section{Fidelity of GHZ-state \label{section-trans-fidelity}}

\begin{figure}[t]\centering
	\subfloat[]{\centering
		\includegraphics[width=0.98\columnwidth, height=4cm]{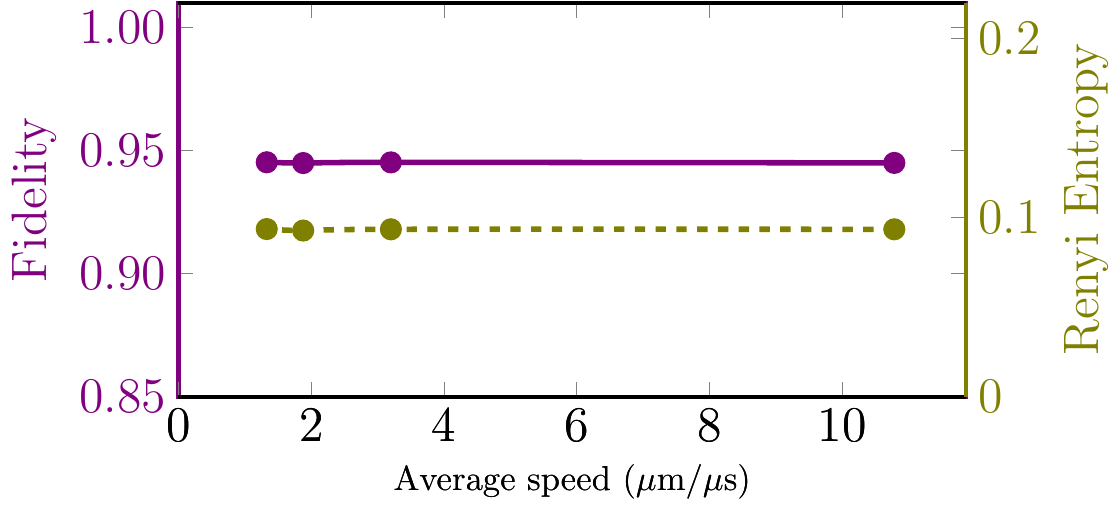}
	}
	\\
	\subfloat[]{
		\includegraphics[width=0.98\columnwidth, height=5cm]{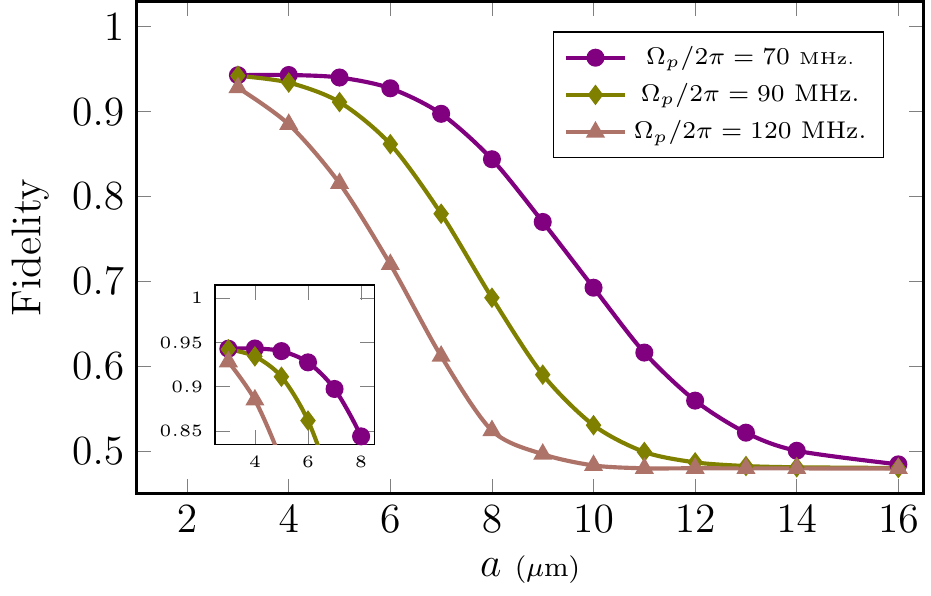}
	}
	\caption{(a) The evolution of fidelity (solid-line), and Renyi entropy (dashed-line) as a function of average speed of transporting the ancilla qubit over a distance $d^{\ast}=d \, \sqrt{2}=60 \, \sqrt{2}=84,85$ \si{\mu m} between non-neighboring data qubits. (c) The evolution of fidelity (a) and Renyi entropy (b) as a function of the minimum distance between the ancilla and the correspondent data qubit for different values of $\Omega_{p}$, with $\Omega_{c}=2.5~\Omega_{p}$, other parameters are the same as in Fig.~\ref{Fig3:truthtable}. }
	\label{Fig5:averagespeedfidelity1}
\end{figure}

We endorsed the definition of fidelity between arbitrary states of a quantum system from \cite{nielsen2000quantum} 
$$\mathcal{F}(\hat{\rho}, \hat{\sigma}) = \text{Tr} \left( \sqrt{ \sqrt{\hat{\rho}} \, \hat{\sigma} \, \sqrt{\hat{\rho}} } \right),$$
where we have considered $\hat{\rho}$ as the calculated density matrix after partially tracing the subspace of logical ground states of control and target atoms, and $\hat{\sigma}=| \Phi^{+} \rangle \langle \Phi^{+} |$ is the density matrix of the multi-qubit entangled state  $|\Phi^{+}\rangle=\frac{1}{\sqrt{2}} \left( \otimes_{\ell}^{N} | 0 \rangle_{\ell} + \otimes_{\ell}^{N} | 1 \rangle_{\ell} \right)$. We calculated the density operator of the system when it was initially prepared in the superposition of the clock states of control atom, which results of applying Hadamard gate initially to the ground state $|0\rangle$. 

We have $N=4$ target atoms, which represent a Greenberger - Horne - Zeilinger (GHZ) state, which is an advantageous resource in quantum computing and cryptography \cite{hillery1999quantum}. 

In Fig.~\ref{Fig5:averagespeedfidelity1}(a), we plot the fidelity as a function of average speed of transporting the ancilla. It is clearly shown that fidelity is not affected by the average speed of moving of the control atom since the transport/separation process occurs while control atom is in the ground state which meets with an experimental work of Bluvstein et al. \cite{bluvstein2022quantum}. In their findings, fidelity is only affected by atom loss as a dominant error mechanism for average speed $>0.55$~\si{\mu m/\mu s}. In Fig.~\ref{Fig5:averagespeedfidelity1}(b), we show the dependence of fidelity on the minimum interatomic distance between control and the corresponding target atom for two different values of $\Omega_{p}=2 \pi \times 70$~\si{MHz}, $2 \pi \times 90$~\si{MHz}, and $2 \pi \times 120$~\si{MHz}. The maximum value of obtained fidelity is $94.96\%$. Fidelity declines by enlarging the distance $a$~(\si{\mu m}). Moreover, the declining rate is higher by increasing the value of $\Omega_{p}$ (considering faster implementation of each CNOT gate).


\section{Renyi entropy and Mutual information\label{section-mutual-information}}

Detection and measurement of entanglement is a fundamental property of quantum systems. If the state function $\psi_{\mathds{A}\mathds{B}}$ of a quantum system is a product state of subsystems $\mathds{A}$, and $\mathds{B}$ of the many body system $\mathds{A}\mathds{B}$, then $$\text{Tr}(\rho_{\mathds{A}}^{2})=\text{Tr}(\rho_{\mathds{B}}^{2})=\text{Tr}(\rho_{\mathds{A}\mathds{B}}^{2})=1,$$ where 	$\rho_{\mathds{A}}=\text{Tr}_{\mathds{B}}(\rho_{\mathds{A}\mathds{B}})$ is the reduced density matrix of subsystem $\mathds{A}$. The tracing over a subsystem indicates ignoring all information about this subsystem. For an entangled state function, the subsystems are less pure compared to the whole system resulting $\text{Tr}(\rho_{\mathds{A}}^{2})<\text{Tr}(\rho_{\mathds{A}\mathds{B}}^{2})$ and $\text{Tr}(\rho_{\mathds{B}}^{2})<\text{Tr}(\rho_{\mathds{A}\mathds{B}}^{2})$. These inequalities can be framed in terms of quantities of quantum entropies, particularly Renyi entropy \cite{horodecki2009quantum}. The $n$-th order Renyi entropy of the subsystem $\mathds{A}$ is given by $$S_{n}(\mathds{A})=\frac{1}{1-n} \log_2 \left( \, \text{Tr} \rho_{\mathds{A}}^{n} \, \right).$$ 

As a limiting case for $n\rightarrow1$, we get von Neumann entropy. For $n=2$, we get the second order Renyi entropy $$S_{2}(\mathds{A})=-\log_2 \left( \text{Tr} \rho_{\mathds{A}}^{2} \right),$$ which is related to purity, providing a lower bound for the von Neumann entanglement entropy. $S_{2}(\mathds{A})$ provides more information about the quantum state than von Neumann entropy, where the sufficient conditions for entanglement to be demonstrated become $S_{2}(\mathds{A})>S_{2}(\mathds{A}\mathds{B})$ and $S_{2}(\mathds{B})>S_{2}(\mathds{A}\mathds{B})$ \cite{islam2015measuring}.

In Fig.~\ref{Fig5:averagespeedfidelity1}(a), we show the evolution of Renyi entropy $S_{\mathds{A B}}$ as a function of the average separation speed of transporting the control atom over a distance ($d^{\ast}-2~a$)~\si{\mu m}, and the minimum distance between control and target atoms $a$, respectively. In Fig.~\ref{Fig5:averagespeedfidelity1}(a), it is clearly shown that Renyi entropy is not affected by change in average speed. Atom loss during the coherent transport process is a potential dominant error mechanism that may affect Renyi entropy. 


\begin{figure}[t]\centering
	\subfloat[$\Omega_{p}=2\pi\times70$ \si{MHz}.]{
		\includegraphics[width=0.5\columnwidth,height=4.5cm]{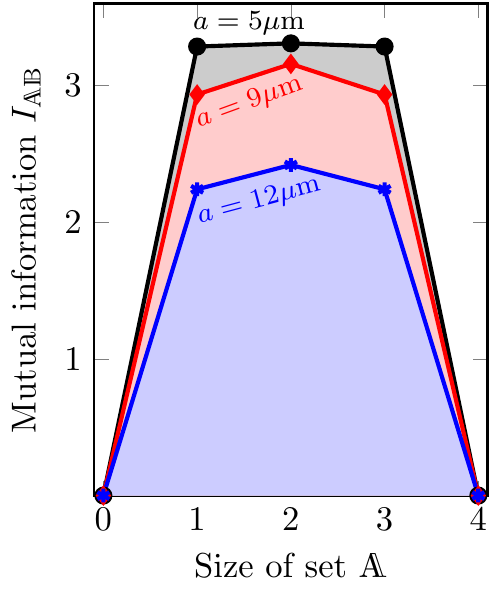}
	}
	\subfloat[$\Omega_{p}=2\pi\times90$ \si{MHz}.]{
		\includegraphics[width=0.45\columnwidth,height=4.5cm]{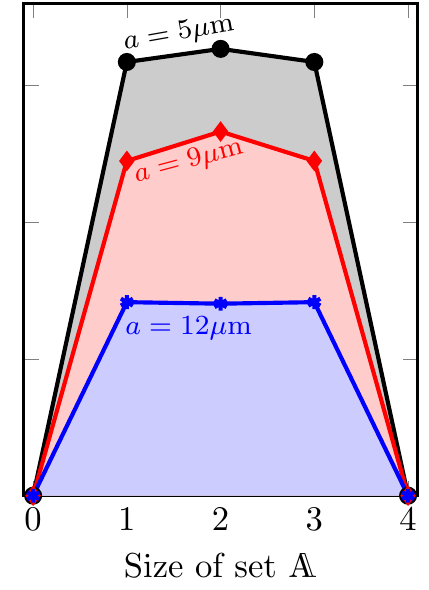}
	}
	\caption{The calculated mutual information $I_{\mathds{AB}}$ as a function of the subset $\mathds{A}$ for different values of the minimum distance between control and target atoms $a$ (\si{\mu m}) and maximum value of Raman pulse amplitude $\Omega_{p}$.}
	\label{Fig6:mutualinformationsizeA}
\end{figure}


Mutual information of two random variables is a measure for the dependence of one of the variables on the other one or the amount of information that can be obtained about one variable by just observing the other variable. It can be understood as a measure of correlation between two variables \cite{wolf2008area}. Renyi mutual information $I_{\mathds{A}\mathds{B}}$ for two sets of atoms $\mathds{A}$ and $\mathds{B}$ is given by 
\begin{equation}\label{eq.mutual.information}
	I_{\mathds{A}\mathds{B}}=S_{2}(\mathds{A})+S_{2}(\mathds{B})-S_{2}(\mathds{A}\mathds{B}),
\end{equation}
where the set $\mathds{A}$ is considered to include the reduced density operator for the control atom with any number of $N$ target atoms and the set $\mathds{B}=\mathds{A}^{c}$. The special case where the set $\mathds{A}$ is empty set or include all target atoms, it is clear that from eq.(\ref{eq.mutual.information}) the mutual information $I_{\mathds{A}\mathds{B}}$ is zero. If the mutual information grows linearly in the system size, then the rate of measurement is low (known as \textit{volume-law}). While if the mutual information does not grow with the system size, then a high rate of measurement prevents entanglement from accumulating in the system (known as \textit{area-law}).

In Fig.~\ref{Fig6:mutualinformationsizeA}, we show the evolution of the mutual information $I_{\mathds{A}\mathds{B}}$ as a function of the size of set $\mathds{A}$, for different values of the minimum distance between control and target atoms $a$~(\si{\mu m}). Since the number of target atoms is limited in the considered architecture $N=4$, it is quite unclear describing the scaling behavior of the system whether revealing \textit{volume} or \textit{area} law. Also, the amount of mutual information decreasing by setting the control atom minimally far from the target atoms and becomes less as in Fig.~\ref{Fig6:mutualinformationsizeA}(b) considering faster implementation of CNOT gate by decreasing the Raman pulse time $T_{p} \sim \frac{1}{\Omega_{p}}$. 

\section{Conclusion\label{section-conclusion}}

In this article, we proposed a 2D scalable heteronuclear architecture for implementing CNOT$^{4}$ gate sequentially by trapping ancillas (Cs atoms) in a trap generated by AOD and data qubits (Rb atoms) in a trap generated by SLM. By coherently transporting the array of ancillas. Using coherent transport, the intraspecies interaction is suppressed since target atoms are set to be far enough from each other. Fidelity of CNOT gates near $\simeq 95\%$ in the case of transporting an ancilla between four different data qubits with $3$~\si{\mu m} is the minimum interatomic distance between ancilla and data qubits, and for moderate Rabi frequencies of Raman laser pulse of $\Omega_{p}=2\pi\times70$~\si{MHz}.  We have shown that the average speed of atomic motion does not affect the dynamics of the qubit states (atomic loss is not included in the model). The mutual information is decreasing by placing the control atom far from the fixed position of target atoms when implementing the gate. Also, a moderate value of the gate speed affects the value of mutual information.

\begin{acknowledgments}
	
This work is supported by the Russian Science Foundation (Grant No.~\href{https://rscf.ru/project/23-42-00031/}{23-42-00031}). A. Farouk acknowledges funding support from the joint executive program between Egypt and Russia (EGY-6544/19). P. Xu acknowledges funding support from the National Key Research and Development Program of China (Grant No. 2021YFA1402001), the Youth Innovation Promotion Association CAS No. Y2021091.

\end{acknowledgments}



\nocite{*}

\bibliography{References,FootNotes}

\end{document}